\documentstyle[12pt]{article}
\input psfig
\def\reference{\parskip 0pt\par\noindent\hangindent 0.5 truecm}

\def\gs{\mathrel{\raise0.35ex\hbox{$\scriptstyle >$}\kern-0.6em 
les
\lower0.40ex\hbox{{$\scriptstyle \sim$}}}}
\def\ls{\mathrel{\raise0.35ex\hbox{$\scriptstyle <$}\kern-0.6em 
ggles
\lower0.40ex\hbox{{$\scriptstyle \sim$}}}}
\def\ltorder{
\mathrel{\raise.3ex\hbox{$<$}\mkern-14mu\lower0.6ex\hbox{$\sim$}}
}
\def\gtorder{
\mathrel{\raise.3ex\hbox{$>$}\mkern-14mu\lower0.6ex\hbox{$\sim$}}
}

\def\ep{\epsilon}

\begin{document}
%
%
\title{Do angular momentum induced ellipticity correlations contaminate 
weak lensing measurements?}
%


\author{Priyamvada Natarajan $^{1}$, Robert G. Crittenden$^{2}$, 
Ue-Li Pen$^{3}$ \& Tom Theuns$^{4}$
} 

\date{}
\maketitle

{\center
$^1$ Department of Astronomy, Yale University, New Haven, CT, USA\\
priya@astro.yale.edu\\[1mm]
$^2$ Department of Applied Mathematics and Theoretical Physics, 
Wilberforce Road, Cambridge CB3 0WA, UK
\\R.G.Crittenden@damtp.cam.ac.uk\\[1mm]
$^3$ CITA, McLennan Labs, University of Toronto, Toronto, M5S 3H8, Canada
\\pen@cita.utoronto.ca\\[1mm]
$^4$ Institute of Astronomy, Madingley Road, Cambridge CB3 0HE, UK\\
tt@ast.cam.ac.uk\\[1mm]
}

%
\begin{abstract}
Alignments in the angular momentum vectors of galaxies can induce
large scale correlations in their projected orientations.  Such
alignments arise from the tidal torques exerted on neighboring
proto-galaxies by the smoothly varying shear field. Weak gravitational
lensing can also induce ellipticity correlations since the images of
neighboring galaxies will be distorted coherently by the intervening
mass distribution. Comparing these two sources of shape correlations,
it is found that for current weak lensing surveys with a median
redshift of $z_m=1$, the intrinsic signal is a contaminant on the
order of 1-10\% of the measured signal. However, for shallower surveys
with $z_m \le 0.3$, the intrinsic correlations dominate over the
lensing signal. The distortions induced by lensing are curl-free,
whereas those resulting from intrinsic alignments are not.  This
difference can be used to disentangle these two sources of ellipticity
correlations. When the distortions are dominated by lensing, as occurs
at high redshifts, the decomposition provides a valuable tool for
understanding properties of the noise and systematic errors.
\end{abstract}

{\bf Keywords:}

\medskip

%
%

\section{Introduction}

Gravitational lensing can be used to map the detailed distribution of
matter in the Universe over a range of scales (Gunn 1967). Systematic
distortions in the shapes and orientations of high redshift background
galaxies (weak lensing) induced by mass inhomogeneities along the line
of sight can be measured statistically (Gunn 1967; Blandford et
al. 1991; Miralda-Escude 1991; Kaiser 1992; see a recent review by
Bartelmann \& Schneider 1999).

The lensing effect depends only on the projected surface mass density
and is independent of the luminosity or the dynamical state of the
mass distribution. Thus, this technique can potentially provide
invaluable constraints on the distribution of matter in the Universe
and the underlying cosmological model (Bernardeau, van Waerbeke \&
Mellier 1997).  There has been considerable progress in theoretical
calculations of the effects of weak lensing by large-scale structure,
both analytically and using ray-tracing through cosmological N-body
simulations (Kaiser 1992; Bernardeau, van Waerbeke \& Mellier 1997;
Jain \& Seljak 1997; Jain, Seljak \& White 2000).

Recently, several teams have also reported observational detections of
`cosmic shear' -- weak lensing on scales ranging from an arc-minute to
ten arc-minutes (see Fig. 1; Van Waerbeke et al. 2000; Bacon,
Refregier \& Ellis 2000; Wittman et al. 2000; Kaiser, Wilson \&
Luppino 2000). At present, these studies are limited by observational
effects, such as shot noise due to the finite number of galaxies and
the accuracy with which shapes can actually be measured given the
optics and seeing (Kaiser 1995; Bartelmann \& Schneider 1999).  In
addition, the intrinsic ellipticity distribution of galaxies and their
redshift distribution is still somewhat uncertain.  These
observational difficulties can be potentially overcome with more data.

However, an important theoretical issue remains. In modeling the
distortion produced by lensing, it is assumed that the {\it a priori}
intrinsic correlations in the shapes and orientations of background
galaxies are negligible. Correlations in the intrinsic ellipticities
of neighboring galaxies are expected to arise from the galaxy
formation process, for example as a consequence of correlations
between the angular momenta of galaxies when they assemble.  The
strength of these correlations can be computed in linear theory, in
the context of Gaussian initial fluctuations.

\section{Schematic Outline}

We briefly outline the calculation here, details can be found in the
following two papers Crittenden et al. (2001a) and Crittenden et
al. (2001b).  To estimate the strength of intrinsic ellipticity
correlations, we approximate the projected shape of a galaxy on the
sky by an ellipsoid with semi-axes $a$, $b$ ($a > b$).  The
orientation of the ellipsoid is given by the angle $\psi$ between the
major axis and the chosen coordinate system, while its magnitude is
given by $|\ep| = (a^2 - b^2)/(a^2 + b^2)$. Both the magnitude of the
ellipticity and its orientation can be concisely described by the
complex quantity $\epsilon^{(o)}$,
\begin{eqnarray} 
\epsilon^{(o)} = |\epsilon^{(o)}| e^{2i\psi} 
= [\epsilon_{+}^{(o)} + i \epsilon_{\times}^{(o)}].
\end{eqnarray} 
where the superscript $^{(o)}$ denotes the 
observed shape. 

In the linear regime and under the assumption of weak lensing, the
lensing equation can be written as,
\begin{eqnarray}
\epsilon^{(o)}\,=\,\frac{\epsilon\,+\,g}{1\,+\,g^{*}\epsilon},
\end{eqnarray}
where $g$ is the complex shear and $\epsilon$ the intrinsic shape of
the source (Kochanek 1990; Miralda-Escude 1991). Furthermore, in the
weak regime, correlations of this distortion field are
\begin{eqnarray}
\langle{\epsilon^{(o)}}(\mathbf{x_1})\,{\epsilon^{(o)*}}(\mathbf{x_2})
\rangle\,\simeq\,
\langle{\epsilon (\mathbf{x_1})}\,{\epsilon^{*}}(\mathbf{x_2})\rangle\,+
\,\langle{g^{*}}(\mathbf{x_2}){\epsilon
(\mathbf{x_1})}+{g(\mathbf{x_1})}
{\epsilon^{*}}(\mathbf{x_2})\rangle
\,+\,\langle{g(\mathbf{x_1})}{g^{*}}(\mathbf{x_2})\rangle\,
\end{eqnarray}
where the $^*$ denotes complex conjugation.  The first term is the
contribution that arises from intrinsic shape correlations. Previous
analyses have focused on the third term of this expression,
correlations due to weak lensing.

We assume in the calculation that shape correlations arise primarily
from correlations in the direction of the angular momentum vectors of
neighboring galaxies.  Spiral galaxies are disk-like with the angular
momentum vector perpendicular to the plane of the disk, so that
angular momentum couplings will be translated into shape correlations.
We will assume that for ellipticals the angular momentum vector also
lies along its shortest axis on average, as it does for the spirals.
However, since elliptical galaxies are intrinsically rounder, the
correlation amplitude will be smaller.  We use the observed
ellipticity distributions of each morphological type (from the APM
survey) in the computation of the shape correlations.  For weak
lensing, in contrast, the induced shape correlations are independent
of the original shapes of the lensed galaxies.

\begin{figure}[h]
\centerline{\psfig{file=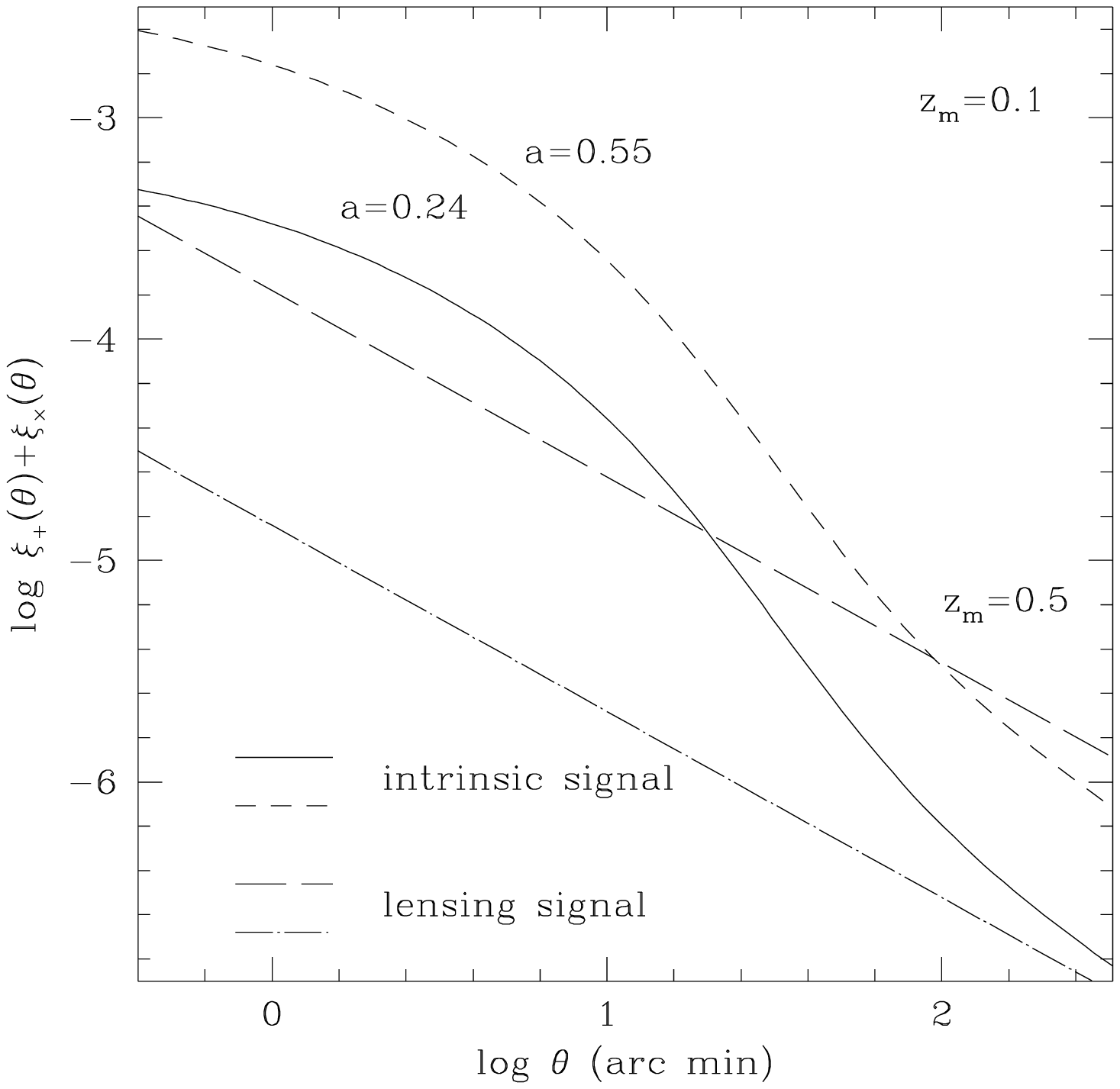,height=2.8in}
\psfig{file=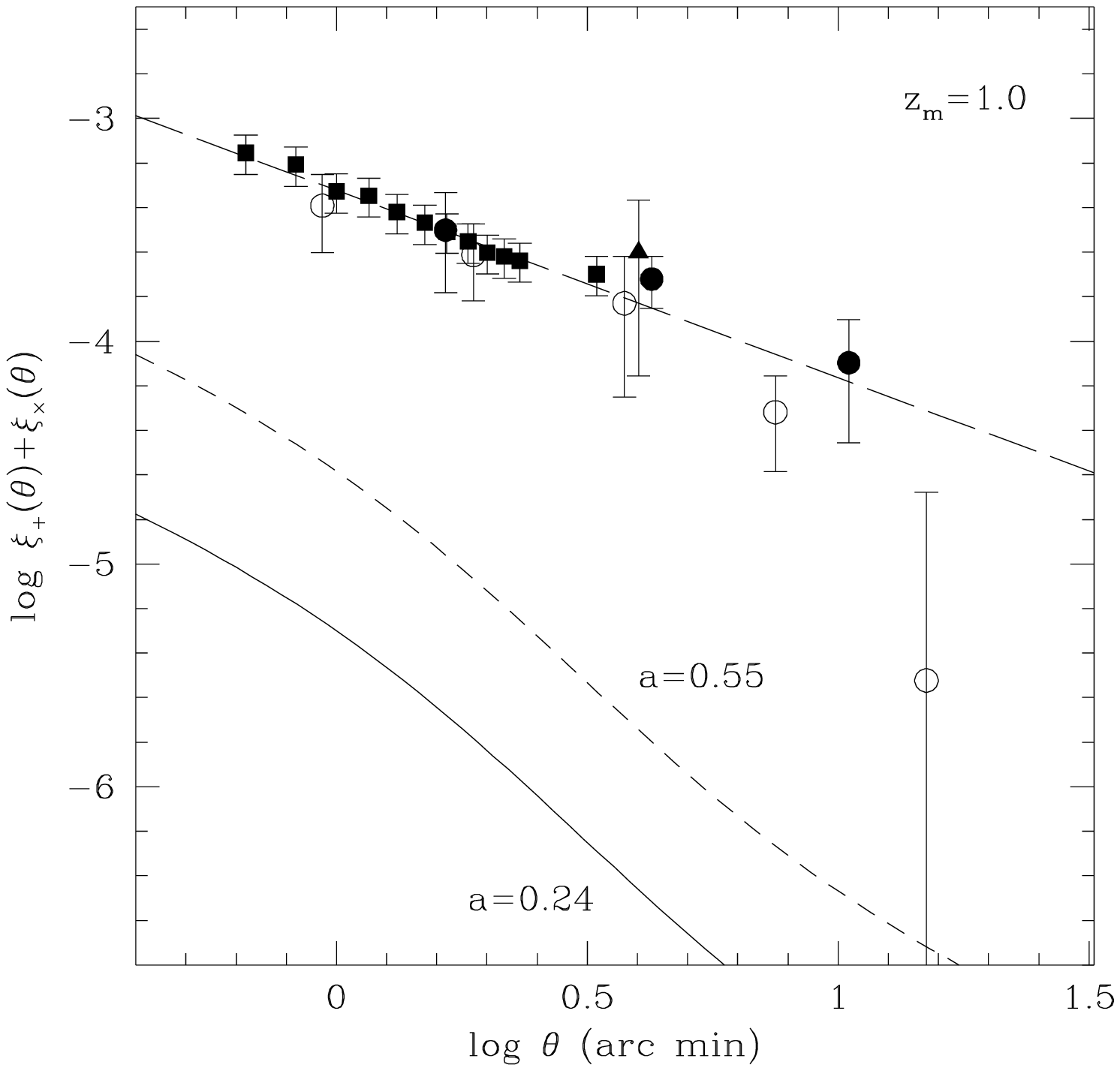,height=2.8in}}
\caption{The intrinsic correlation signal versus the predictions from
weak lensing and current observations. Right panel:
$\xi_+(\theta)+\xi_\times(\theta)$ the intrinsic signal for $z_m = 1$, 
compared to the measured shear correlation function. 
At small separations, the intrinsic signal is approximately 
1\% of the lensing signal. The amplitude depends on the 
assumed average galaxy thickness ($\alpha$) and the 
parameter $a$ that describes how well the angular momentum of the 
galaxy is correlated with the shear field. We plot $a=0.24$ 
(full line) and $a=0.55$ (short-dashed line) which correspond 
to the values inferred from numerical simulations.
$\alpha=0.73$ corresponds to the value determined from the observed
distribution of ellipticities. The data are: van
Waerbeke et al. (2000) -- solid squares ; Wittman et al. (2000) --
filled circles ; Kaiser et al. (2000) -- open circles; and Bacon et
al. (2000) -- filled triangle. The long-dashed line is the theoretical
prediction from Jain \& Seljak (1997) computed for a
$\Omega_\Lambda=0.7$ galaxy cluster normalized flat universe, $\sim
4.75\times10^{-4}(\theta/{\rm arc min})^{-0.84}$. Left panel:
predictions for a shallower survey such as SDSS and 2dF with 
$z_m=0.1$. The intrinsic signal is plotted for 2 values
of $a$, and the theoretical prediction for weak lensing is the
long-dashed line (for $z_m=0.1$) and dotted-long-dashed (for
$z_m=0.5$). The lensing prediction for $z_m=0.1$ is extrapolated from
the Jain \& Seljak fit beyond the stated range of validity. For such
low redshifts the intrinsic signal dominates on most scales.}
\end{figure}

\section{Results}

The amplitude and shape of the computed ellipticity correlation
function can be understood intuitively. The ellipticity is a function
of the shear tensor, which is the second derivative of the
potential. By virtue of Poisson's equation, the trace of the shear
tensor is the density.  Therefore, we expect the correlation of the
other components of the shear field will drop at the same rate as the
density correlation function. Since the ellipticities are quadratic in
the shear field, correlations in them will fall as the density
correlation function {\it squared}, $\langle \epsilon
(\mathbf{x_1})\epsilon^{*}(\mathbf{x_2})\rangle \propto \xi_{\rho}^2$
(see Fig. 1).

Comparing the strength of the intrinsic correlation to that expected
for weak lensing, we find that the intrinsic signal {\bf{grows}} as the
depth of the survey {\bf{decreases}} (the projected intrinsic shape 
correlation function scales as $z_m^{-2}$ whereas the weak lensing 
correlation function scales as $z_m^{1.52}$), because in that case galaxies
close on the sky are also on average physically closer, and are hence
more correlated. The weak lensing signal, on the other hand, drops
off, since typically there is less matter between us and the lensed
objects.  For typical weak lensing surveys, however, with a median
redshift of $z_m=1$, the intrinsic signal is between 1 -- 10 per cent
of the weak lensing amplitude. For shallower surveys such as SDSS or
2dF, the intrinsic signal may dominate the lensing one, on small
scales (see Left panel of Fig. 1). Therefore, SDSS and 2dF are ideally
suited for studying intrinsic correlations in the orientations of
galaxies.

The intrinsic ellipticity depends on the square of the tidal field,
whereas the lensing distortion is linear in the shear. As a direct
consequence, the distortion field is curl-free when induced by
lensing, but not when intrinsic correlations are present as well
(Crittenden et al. 2000b). Angular momentum couplings produce $E$ and
$B$-modes in comparable amounts and one might expect that noise,
telescope distortions and other sources of systematic errors will
produce curl modes as well. The detection of such \lq magnetic\rq~
modes will be an invaluable way of separating lensing from intrinsic
correlations. Details on how to unambiguously do so are presented in
Crittenden et al. (2001b).

\section*{Acknowledgements}

Rachel Webster and the LOC are thanked for organizing an excellent, 
interactive and fun workshop.


\section*{References}






\reference{} Bacon, D., Refregier, A. \& Ellis, R. S., 2000, MNRAS, 318, 625
\reference{} Bardeen, J., Bond, J. R., Kaiser, N., \& Szalay, A., 1986, ApJ, 304, 15
\reference{} Bartelmann, M., \& Schneider, P., 1999, Review for Physics Reports, preprint, astro-ph/9909155 
\reference{} Bernardeau, F., van Waerbeke, L., \& Mellier, Y., 1997, A\&A, 322, 1
\reference{} Blandford, R. D., Saust, A. B., Brainerd, T. G., \& Villumsen, J. V.,
1991, MNRAS, 251, 600
\reference{} Catelan, P., \& Theuns, T., 1996, MNRAS, 282, 436 
\reference{} Catelan, P., Kamionkowski, M., \& Blandford, R. D., 
2001, MNRAS, 320, L7
\reference{} Crittenden, R., Natarajan, P., Pen, U. \& Theuns, T., 2000a, 
ApJ submitted, astro-ph/0009052.
\reference{} Crittenden, R., Natarajan, P., Pen, U. \& Theuns, T., 2000b, 
ApJ submitted, astro-ph/0012336.
\reference{} Croft, R. A. C., \& Metzler, C., 2000, preprint, astro-ph/0005384 
\reference{} Gunn, J., 1967, ApJ, 150, 737
\reference{} Heavens, A., Refregier, A., \& Heymans, C., 2000, MNRAS, 319, 649
\reference{} Jain, B., \& Seljak, U., 1997, ApJ, 484, 560
\reference{} Jain, B., Seljak, U., \& White, S. D. M., 2000, ApJ, 530, 547 
\reference{} Kaiser, N., 1992, ApJ, 388, 272  
\reference{} Kaiser, N., Wilson, G., \& Luppino, G., 2000, preprint,
astro-ph/0003338
\reference{} Lee, J., \& Pen, U., 2000, ApJ, 532, L5 
\reference{} Lee, J., \& Pen, U., 2000, astro-ph/0008135
\reference{} Miralda-Escude, J., 1991, ApJ, 380, 1
\reference{} Pen, U., Lee, J., \& Seljak, U., 2000, preprint, astro-ph/0006118
\reference{} Wittman, D., Tyson, J. A., Kirkman, D., Dell'Antonio, I, \& 
Bernstein, G., 2000, Nat, 405, 143


\end{document}